\documentclass[sigconf]{acmart}

\AtBeginDocument{%
  }

\setcopyright{acmlicensed}
\copyrightyear{2025}
\acmYear{2025}
\acmDOI{XXXXXXX.XXXXXXX}

\acmConference[CHI '25]{ACM CHI Conference on Human Factors in Computing Systems}{April 26--May 1,
  2025}{Yokohama, Japan}
\acmISBN{978-1-4503-XXXX-X/18/06}

\newcommand{\tocite}[1]{{\color{red} [TO CITE]}}

\begin{document}

\title{Mindalogue: LLM‑Powered Nonlinear Interaction for Effective Learning and Task Exploration}



\author{Rui Zhang}
\authornote{Both authors contributed equally to this research.}
\email{gloriezhang777@gmail.com}
\affiliation{%
  \institution{The Future Laboratory, Tsinghua University}
  \city{Beijing}
  \country{China}
}

\author{Ziyao Zhang}
\authornotemark[1]
\email{zhanz098@newschool.edu}
\affiliation{%
  \institution{The New School, New York City}
  \city{New York}
  \country{United States}
}

\author{Fengliang Zhu}
\email{3220231275@bit.edu.cn}
\affiliation{%
  \institution{Beijing Institute of Technologyy}
  \city{Beijing}
  \country{China}
}

\author{Jiajie Zhou}
\email{3j1227713500@gmail.com}
\affiliation{%
  \institution{Beijing Institute of Technologyy}
  \city{Beijing}
  \country{China}
}

\author{Anyi Rao}
\authornote{Corresponding Author.}
\email{anyirao@ust.hk}
\affiliation{%
  \institution{Art and Machine Creativity, Hong Kong University of Science and Technology}
  \city{Hong Kong}
  \country{China}
}

\renewcommand{\shortauthors}{Zhang et al.}
\begin{abstract}
  Current generative AI models like ChatGPT, Claude, and Gemini are widely used for knowledge dissemination, task decomposition, and creative thinking. However, their linear interaction methods often force users to repeatedly compare and copy contextual information when handling complex tasks, increasing cognitive load and operational costs. Moreover, the ambiguity in model responses requires users to refine and simplify the information further.
  To address these issues, we developed "Mindalogue", a system using a non-linear interaction model based on "nodes + canvas" to enhance user efficiency and freedom while generating structured responses. A formative study with 11 users informed the design of Mindalogue, which was then evaluated through a study with 16 participants. The results showed that Mindalogue significantly reduced task steps and improved users’ comprehension of complex information. This study highlights the potential of non-linear interaction in improving AI tool efficiency and user experience in the HCI field.
\end{abstract}




\begin{teaserfigure}
  \includegraphics[width=\textwidth]{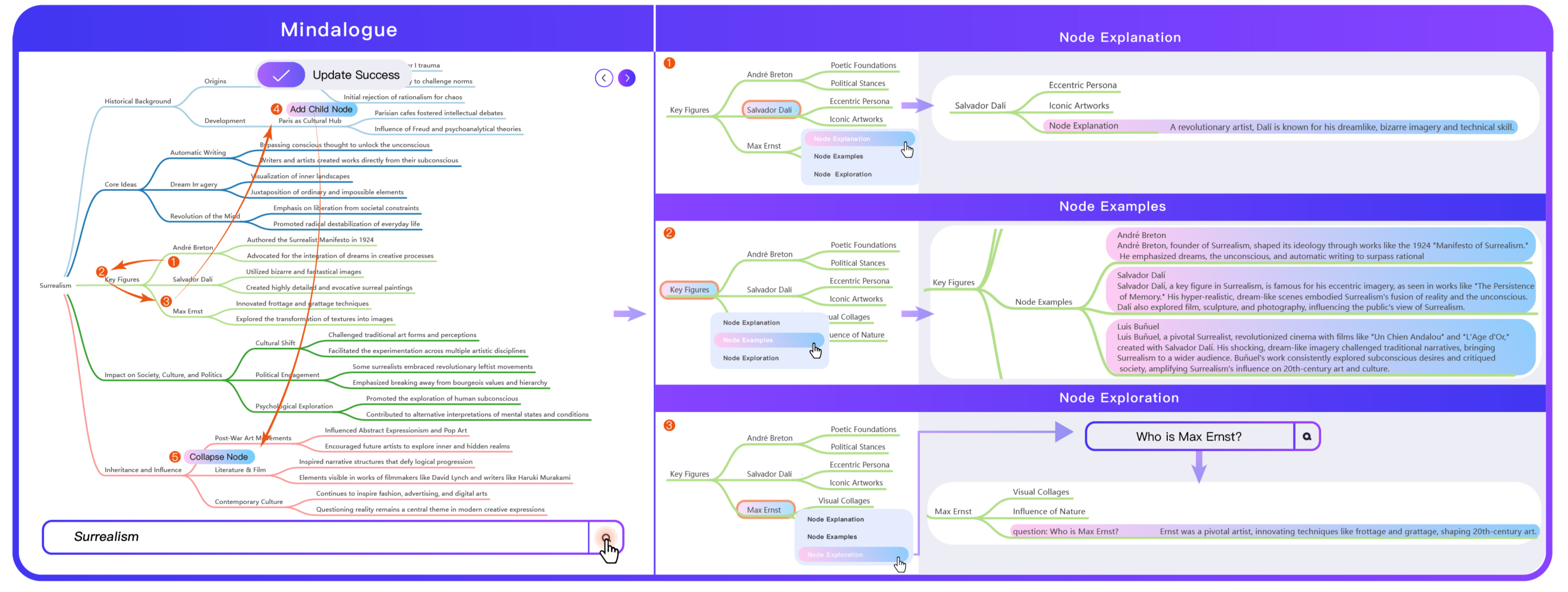}
  \vspace{-20pt}
  \caption{Mindalogue system's non-linear interaction interface, illustrating the flexible connections between nodes and the canvas.}
  \Description{.}
  \label{fig:teaser}
  \Description{This image provides an overview of the system functions. On the left, it displays the overall operation process and the resulting mindmap, showing how users generate the final mindmap by interacting with the nodes. On the right, the system's functional operations and results are presented, detailing the outcomes of the "Node Explanation," "Node Example," and "Node Exploration" features. The purpose of this image is to demonstrate the complete operation flow of the system and how each function influences the final result.}
\end{teaserfigure}

\maketitle

\section{Introduction}

Large Language Models (LLMs) such as ChatGPT~\cite{ChatGPT2023}, Claude~\cite{Claude2023}, Gemini~\cite{Gemini2023} generate high-quality text from natural language prompts and are used in various fields such as task decomposition ~\cite{Wei2022Chain}, creative thinking ~\cite{Mirowski2022CoWriting}, writing ~\cite{Brown2020Language}, coding ~\cite{Chen2021Evaluating}, and data analysis ~\cite{Bubeck2023Sparks}. They primarily rely on text-based linear interaction~\cite{Gao2019NeuralApproaches}. In this format, users pose a question, and the system generates a response based on that query. Subsequent questions depend on the previous responses, forcing users to proceed step by step without the ability to skip steps or content. This creates a strong dependency between each question and answer, resulting in a linear interaction path. This format is effective for tasks requiring step-by-step execution or scenarios where information needs to be conveyed linearly, such as Q\&A, dialogue practice, and legal analysis~\cite{McTear2020ConversationalAI}.


Nevertheless, when dealing with complex tasks with large amounts of information, multiple layers, and many subtasks—such as brainstorming ~\cite{Mirowski2022CoWriting}, structured knowledge learning ~\cite{Sweller2011CognitiveLoad}, and large project analysis—users need to grasp the system's logic and switch between subtasks for exploration. The linear interaction in current LLMs does not allow for flexible exploration, forcing users to repeatedly compare, modify, and copy previous content. This lowers interaction efficiency and increases cognitive load ~\cite{Sweller2011CognitiveLoad}. Additionally, the outline or paragraph-based responses are not ideal for quickly identifying key information. In multi-turn conversations, useful information is often scattered ~\cite{Kocielnik2018ReflectiveConversations}, with overlaps or conflicts in the overall answer. This unstructured outcome limits the user's ability to deeply understand and organize the content.

To address these issues, we conducted a formative study with 11 participants to identify challenges they faced when using LLMs for complex tasks. These included difficulties with long-context conversations, handling multiple subtasks, lengthy responses, and the information retrieval challenges and cognitive load caused by linear interaction systems. Based on these insights, we designed an LLM-based interactive system called Mindalogue. This system uses a "node + canvas" approach, with a mindmap structure that helps users clearly understand the logical relationships between different parts of a task or system. The node-based input-output and non-linear interaction offer users greater flexibility and freedom in their interactions with the system.

In our evaluation study, we examined how 16 users used Mindalogue and traditional LLMs(ChatGPT) to learn or explore tasks on both familiar and unfamiliar topics. We compared the advantages and limitations of both systems in terms of interaction style and generated outcomes. We also investigated how each system impacted users' operational freedom and whether structured outcomes could reduce cognitive load and increase reliability in the results. 

In general, our work contributes to the following aspects.
\begin{itemize}
\item \textbf{A formative study} with 11 participants to collect and analyze feedback and needs when using current LLMs. This study revealed the limitations of LLMs in performing complex tasks using linear interaction and response structures.

\item \textbf{The "Mindalogue" prototype}, which employs a non-linear interaction approach to generate graphically structured answers. It provides features for single-node explanations (explain), examples (examples), and exploration (explore), allowing users to conduct multi-level AI explorations based on topics and freely delete, add, edit or move node content. 

\item \textbf{An evaluation study} with 16 participants to compare the Mindalogue system with traditional linear interaction LLM products, we aim to validate Mindalogue's improvements in exploration space, user freedom, and the enhancement of user cognition and reliability in outcomes through visualized structured results.
\end{itemize}

\section{Related Work}
This research aims to enhance user flexibility and comprehension efficiency when interacting with LLMs through the design of a non-linear interaction system. To achieve this, we reviewed literature on linear vs. non-linear interaction, text visualization and structuring techniques, and cognitive frameworks for users engaging in complex tasks. Below, we review prior work across these three areas.

\subsection{Linear Interaction vs. Non-linear Interaction}
Linear interaction models have long dominated human-computer interaction design, particularly in command-line interfaces and early dialogue systems~\cite{weizenbaum1983eliza,winograd1972understanding}. User inputs and system responses follow a sequential progression, which is effective for single-task execution by simplifying operations and reducing cognitive load~\cite{norman1991cognitive}. For instance, tasks are completed by following fixed steps, reducing the need for complex cognitive processing~\cite{card2018psychology}. However, when handling multi-layered tasks, users must frequently compare and replicate contextual information, which increases cognitive load and limits exploration flexibility, leading to operational redundancy~\cite{shneiderman1983direct}.

\begin{figure}[t]
  \centering
  \includegraphics[width=\linewidth]{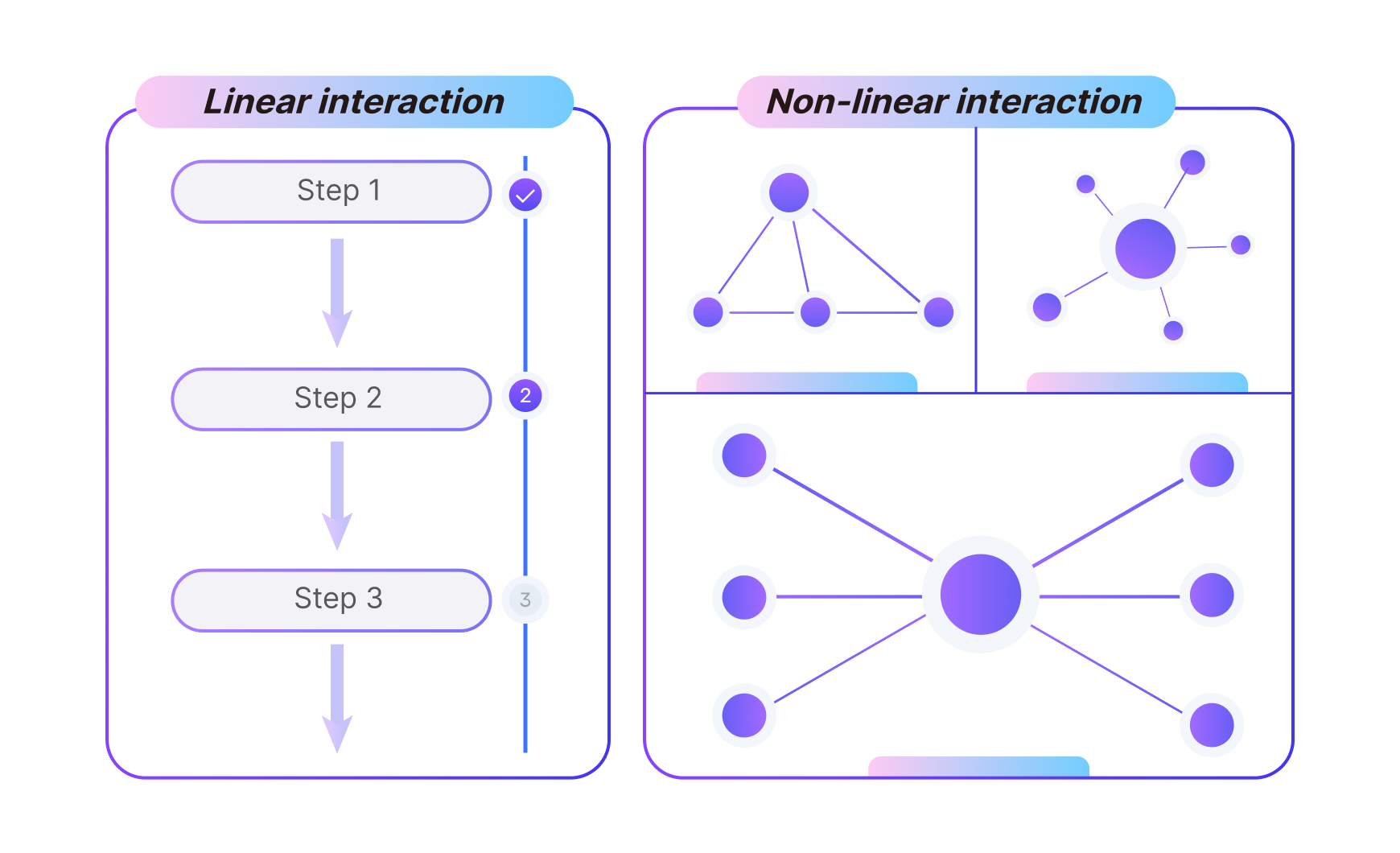}
    \vspace{-10pt}
  \caption{Linear Interaction vs Non-linear Interaction. Linear interaction following a fixed sequence, while non-linear interaction allows flexible exploration and dynamic content structuring through a network of nodes.}
  \label{Linear Interaction vs Non-linear Interaction}
  \Description{This is an interaction mode distinction diagram, showing the difference between 'linear interaction' and 'Non-linear interaction'. On the left are three steps of linear interaction in sequence, and on the right are three different types of nonlinear interactions, showing the free connection between multiple nodes.}
\end{figure}

In contrast, non-linear interaction models provide users with greater exploration space, allowing for more flexible task management and better handling of complex information structures~\cite{johnson1989xerox,holmquist1997focus+}(Figure~\ref{Linear Interaction vs Non-linear Interaction}). For example, Pad++ uses zooming techniques to enable seamless navigation across multi-layered information, reducing the cognitive load associated with frequent context switching~\cite{bederson1994pad++}. Similarly, Fisheye Views dynamically adjust focus, allowing users to view both local details and global information, mitigating cognitive overload~\cite{furnas1986generalized,fairchild2013semnet}. Experiments have shown that fisheye views provide more global context in complex hierarchical structures, significantly improving task completion speed~\cite{schaffer1996navigating}. Additionally, multiscale techniques enhance readability by assigning weights to visual features, avoiding the visual clutter typical in traditional information displays~\cite{baudisch2004multiblending}. The Sensecape system also supports multi-level non-linear interaction, allowing users to flexibly switch between canvas view and hierarchical view, helping them explore and reason across different levels of abstraction. By externalizing complex information, it enables more efficient navigation and understanding~\cite{suh2023sensecape}. This system addresses the mismatch between linear dialogue and complex tasks, further demonstrating the advantages of non-linear interaction.

For information workers, tasks often span multiple domains and require frequent switching between tasks~\cite{gonzalez2004constant}. This fragmented work pattern exacerbates the limitations of linear interaction, making multi-tasking inefficient. non-linear interaction better supports this fragmented work style by offering flexible task management and quick switching mechanisms~\cite{gonzalez2004constant}.

Our Mindalogue system introduces non-linear interaction into LLMs for the first time by incorporating spatial-filling techniques from Tree-Maps, enabling users to structure information through a "nodes + canvas" approach and dynamically adjust node size for flexible navigation of multidimensional data~\cite{johnson1998tree}. This design is similar to the Graphologue system, which converts LLM text responses into graphical diagrams, helping users better understand complex information through node-link diagrams~\cite{jiang2023graphologue}. Graphologue constructs node maps in real-time through prompt strategies and interaction design, fostering non-linear dialogue and information exploration between humans and LLMs, further demonstrating the advantages of non-linear interaction for supporting complex tasks~\cite{jiang2023graphologue}.

Moreover, the radial layout design of MoireGraphs increases efficiency in exploring complex information, allowing users to quickly switch between dimensions by controlling focus intensity~\cite{jankun2003moiregraphs}. Meanwhile, TreePlus uses tree structures and cross-linking techniques to let users incrementally expand complex information, unconstrained by linearity~\cite{lee2006treeplus}. InterRing's multi-focal distortion technology further enhances user flexibility and freedom when handling complex tasks~\cite{yang2002interring}. Research has also shown that non-linear interaction exhibits significant advantages in handling dense information, particularly in path-exploration tasks~\cite{ghoniem2005readability}.

Overall, non-linear interaction models offer greater flexibility and efficiency for processing complex tasks. By integrating these techniques, Mindalogue optimizes the user experience with LLMs, reducing cognitive load and enhancing task handling flexibility and efficiency.

\subsection{Text Visualization and Structuring}

As the volume of information grows, effectively visualizing and structuring complex text becomes a critical challenge. Text visualization techniques help users identify patterns and relationships within data, and are widely applied in data analysis, education, and collaboration~\cite{kucher2015text,chen2010information,ainsworth2003effects,dunne2013motif}. For example, SentenTree uses node-link diagrams to display word co-occurrence relationships, enabling users to quickly grasp key concepts within large text collections~\cite{hu2016visualizing}. Similarly, DocuBurst uses a WordNet-based hierarchical structure for geometric and semantic zooming to represent text semantics~\cite{collins2009docuburst,sebrechts1999visualization}. Meanwhile, ICMAP provides interactive concept maps to help students construct knowledge structures~\cite{fatemeh2011icmap,mcclellan2004cnt}.

In terms of integrating text and data, CrossData automatically generates text-data connections, reducing the manual burden when documenting data~\cite{chen2022crossdata,cui2019text}. Similarly, MUE presents knowledge structures through nested frames, aiding navigation~\cite{travers1989visual}. Pad++ supports multi-level information navigation via a zooming graphical interface~\cite{bederson1994pad++}, similar to booc.io's layered circular layout~\cite{schwab2016booc}. Additionally, PhotoMesa uses quantum treemaps and bubble diagrams to browse large image collections, providing easy navigation~\cite{bederson2001photomesa}.

In designing information visualization, Tufte emphasizes clear and effective chart design to present complex data, reducing visual noise and redundant information~\cite{10.5555/33404,card1999using}. This visual simplification allows the Mindalogue system to present clearer information structures, effectively conveying complex content. Further, Ware, from a perceptual science perspective, explains how optimizing information transmission through visual channels is crucial, particularly in multi-level information layout design~\cite{inbook}.

Mindalogue combines LLM text generation with the "nodes + canvas" layout to transform text into non-linear graphical content, enabling users to explore and reorganize information flexibly~\cite{spoerri1993infocrystal}. Like InfoCrystal, users can adjust content and conduct complex queries~\cite{spoerri1993infocrystal}. ConceptScape synchronizes concept maps with video content, supporting content-based learning~\cite{liu2018conceptscape}.

In document visualization, DocuBurst and SentenTree aid document comparison through lexical hierarchies and co-occurrence relationships~\cite{collins2009docuburst,hu2016visualizing}. ConceptGuide organizes unstructured content using concept maps~\cite{tang2021conceptguide}. Moreover, Log-it uses interactive logging to help users handle complex tasks~\cite{jiang2023log}. To improve the readability of information networks, Dunne and Shneiderman propose motif simplification, replacing common motifs such as fans, connectors, and factions with compact glyphs, enhancing visual clarity~\cite{dunne2013motif}.

In summary, Mindalogue integrates various techniques to offer flexible non-linear visualization and structuring methods, enhancing user exploration, comprehension, and task processing efficiency.

\subsection{User Task Cognition and Exploration}
Managing cognitive load is a core challenge in complex tasks. Cognitive load theory posits that overly complex tasks reduce efficiency~\cite{sweller1988cognitive,wickens2002multiple}. Techniques like progressive disclosure can effectively reduce cognitive load~\cite{van2005cognitive}. The DeFT framework further highlights that multi-representational designs distribute cognitive tasks, optimizing the user’s cognitive experience~\cite{ainsworth2006deft}.

Visual representation is crucial for task cognition. The theory of cognitive fit asserts that task efficiency improves when representation matches the task~\cite{baker2009using,tversky2002animation}. For instance, the Pad++ system uses semantic zooming to help users switch between information layers, reducing cognitive load~\cite{bederson1994pad++}. Similarly, Graphologue enhances user flexibility in complex tasks through graphical nodes~\cite{jiang2023graphologue}.

Flexible task decomposition is also critical. AI Chains supports users in completing complex tasks step-by-step through chained decomposition techniques~\cite{wu2022ai,capra2010tools,zhang2023language}. Mindalogue allows users to dynamically adjust information nodes through its "nodes + canvas" structure, improving task exploration efficiency~\cite{jiang2023graphologue}.

Information-seeking strategies are equally important in task cognition. Information Foraging Theory explains how users adjust information-seeking strategies based on task demands to maximize information retrieval efficiency~\cite{pirolli1999information}. This theory informs Mindalogue's design, enabling users to better organize and filter information through flexible node management, optimizing task cognition and exploration~\cite{jiang2023graphologue}.

In task integration, graphical representation enhances user cognition. Chang et al. found that concept mapping techniques significantly aid users in understanding and summarizing complex information, particularly in multi-level information processing tasks~\cite{chang2002effect}. Hahn and Kim emphasize the importance of flexible graph design for task and information integration~\cite{hahn1999some}. Sedig and Parsons’ cognitive activity support tools (CASTs) utilize multi-representational designs to help users optimize cognitive load in complex reasoning and decision-making tasks~\cite{sedig2013interaction}.

Mindalogue's multi-level flexible structure demonstrates how dynamic node adjustments support continuous discovery of new information and task ideas during exploration. This flexibility parallels the Sparks system's "spark" sentence generation, which inspires user creativity and exploration using language models~\cite{gero2022sparks}.

In conclusion, the Mindalogue system demonstrates excellent adaptability and flexibility in complex task cognition and exploration by integrating cognitive load theory, semantic zooming, information foraging theory, task decomposition, and progressive disclosure strategies~\cite{sweller1988cognitive,bederson1994pad++,jiang2023graphologue,baker2009using,pirolli1999information,wu2022ai,chang2002effect,hahn1999some,ainsworth2006deft,sedig2013interaction,gero2022sparks}.

\section{Formative Study}

To gain an in-depth understanding of the challenges users face with existing linear conversational systems, including operational complexity, increased cognitive load, and insufficient information structuring, we conducted a formative study. This study aims to uncover the primary challenges users encounter in multi-turn interactions with generative AI and provide empirical support for designing nonlinear interaction models.

\subsection{Participants and Procedure}
We interviewed 11 participants (denoted as P1–P11) with varying levels of experience using generative AI, including 3 novice users, 4 moderately experienced users, and 4 advanced daily users.
Participants came from diverse professional backgrounds, such as finance, fast-moving consumer goods, virtual imaging, software engineering, cultural industry management, design research, and mechanical engineering, ensuring a wide range of domain coverage. 
Their ages ranged from 19 to 29 (Mean = 23.91, SD = 2.57), including 6 women and 5 men.

All interviews were conducted remotely and lasted between 45 and 60 minutes. The interviews began by collecting participants’ background information and their experiences with generative AI. They were then asked to reflect on recent long conversations with generative AI, focusing on key challenges and coping strategies in information retrieval, content organization, and logical structuring. Participants also shared specific prompts and actions, revealing their workflows and encountered issues.

The interviews are structured around the following key points:
\begin{itemize}
\item \textbf{Satisfaction with system responses}: Explored participants' overall satisfaction with response speed, accuracy, and logical consistency, especially when handling complex tasks.

\item \textbf{Challenges in maintaining context}: Investigated difficulties participants encountered in maintaining context during long conversations, especially in retrieving information and navigating dialogue across multiple topics and tasks.

\item \textbf{Structuring and applicability of generated content}: Evaluated participants' needs for structured content, its logical coherence, and its applicability to real tasks.
\end{itemize}

\subsubsection{Method}
\paragraph{{Data Collection}}
All interviews were recorded and transcribed to ensure data completeness and accuracy. The interviews focused on participants’ experiences using generative AI systems, particularly challenges encountered during long conversations, multi-task handling, and information structuring.

\paragraph{{Data Analysis}}
We employed reflexive thematic analysis to systematically analyze the transcriptions and identify the major issues participants encountered when interacting with generative AI systems. Three coders independently generated 618 initial codes, which were subsequently discussed, merged, and refined into major themes that provided empirical support for the design of non-linear interaction systems.

\subsection{Findings and Discussion}
Through a detailed analysis of participants’ interaction experiences, we identified the key limitations of current generative AI systems in handling complex tasks and long conversations. Four primary\textbf{D}esign \textbf{C}onsiderations emerged, guiding the future design of non-linear interaction systems.

\paragraph{\textbf{DC1. Limitations and operational inconvenience of linear interaction.}}
Existing linear interaction systems exhibit significant operational inconvenience during long conversations, particularly when handling cross-topic or multi-task operations. For example, participants noted that finding specific information in long conversations required frequent scrolling through conversation logs, which significantly reduced operational efficiency (P6). Additionally, linear interaction relies on a step-by-step - Q\&A model, with each response highly dependent on the preceding question, causing inconvenience when participants asked cross-topic questions (P8). This model limits flexibility and fails to support participants’ needs for efficient operations in complex task scenarios, highlighting the necessity of non-linear interaction models that can better meet participants' requirements for flexible information retrieval and task execution.

\paragraph{\textbf{DC2. Expectation for deeper answers and exploration.}}
Participants expected generative AI to provide more in-depth and targeted responses, especially when dealing with complex issues. For instance, participants noted that system responses were often too broad and lacked sufficient depth to effectively meet the demands of complex tasks (P2, P7). They expressed a desire for systems to iteratively provide responses with more depth and layering. This suggests that future system designs need to enhance AI’s ability to support in-depth exploration, better aligning with participants’ needs in complex tasks.

\paragraph{\textbf{DC3. Need for graphical representation.}}
When handling complex information, participants widely believed that graphical representation could significantly enhance understanding and operational efficiency. Many participants pointed out that purely textual content made it difficult to quickly grasp multi-level information structures, leading to comprehension challenges (P1, P2). Compared to plain text, graphical tools such as mindmaps or flowcharts were considered more intuitive in presenting the organizational structure of information, helping participants make sense of complex content and deepen their understanding. Participants agreed that graphical representation not only improved the readability of information but also significantly enhanced memory retention and operational efficiency. Therefore, future system designs should prioritize converting complex text into intuitive graphical structures to improve task processing efficiency.

\paragraph{\textbf{DC4. Need for structured information and information management.}}
Participants clearly expressed the need for structured information, particularly in long conversations and complex task scenarios. Many participants noted that tables, flowcharts, and mindmaps were helpful in organizing complex information and maintaining logical coherence (P2, P7). Existing generative AI systems fall short in organizing and structuring information, making it difficult for participants to effectively manage and utilize large amounts of information. Future systems should prioritize providing stronger tools for structuring information and logical frameworks, helping participants more efficiently organize, present, and manage information to better support complex task handling.

\subsection{Summary}
Through participants' interviews and data analysis, this study revealed the critical limitations of current generative AI systems in handling complex dialogues and multi-task scenarios. LLM users commonly faced challenges in information retrieval, maintaining context, and the lack of structured generated content, which hindered operational efficiency and user satisfaction. Furthermore, the linear conversational model failed to meet the needs of multi-tasking, limiting system flexibility and adaptability.

To address these challenges, this study provides key empirical support for designing non-linear interaction systems. By incorporating graphical and structured logic, future system designs can better help users handle complex tasks, maintain information coherence, and improve the readability and efficiency of processing information. The ultimate goal is to enhance the user experience through non-linear interaction, making AI-generated content more effective and practical in complex tasks.
\section{Design Goals}
Based on a deep analysis of the limitations of current generative AI systems and users’ needs for system interaction, content generation, and information organization, this study proposes four primary \textbf{D}esign \textbf{G}oals aimed at addressing interaction challenges in complex task scenarios and optimizing system performance.

\paragraph{\textbf{DG1. Provide non-linear interaction to enhance users’ freedom of exploration.}}
Current linear interaction systems exhibit operational inconvenience in long conversations and complex tasks, particularly in cross-subject task switching (\textbf{DC1}). To prevent users from repeating operations and losing information when switching between different subjects, the system should support users in freely exploring during task execution, allowing for quick transitions between sub-topics. Non-linear interaction enables users to seamlessly switch between different topics and tasks while maintaining context consistency, reducing repetitive actions, and ensuring smooth information flow. Specifically, users can utilize a node-based navigation system to flexibly select different task paths, enhancing operational efficiency in complex tasks.

\paragraph{\textbf{DG2. Provide deep information feedback to support in-depth exploration.}}
Users expect more in-depth feedback, particularly when handling complex tasks. Existing systems struggle to provide sufficient detailed support (\textbf{DC2}). Therefore, the system must possess progressive information feedback capabilities, allowing users to further explore the content of a particular node. Through step-by-step feedback mechanisms, users can dive deeper into the details of a topic, gradually transitioning from general information to deeper layers of content. This mechanism guides users in asking more specific questions, with the system continuously offering relevant, in-depth responses, helping users gain a comprehensive understanding of complex topics and satisfying their dual needs for information breadth and depth.

\paragraph{\textbf{DG3. Simplify understanding of complex tasks through graphical representation.}}
When handling multi-dimensional information, plain text struggles to effectively convey the hierarchical structure of complex content, leading to cognitive overload for users (\textbf{DC3}). To address this issue, the system will use graphical tools (such as mindmap) to display task structures, helping users intuitively understand the complex relationships and logical chains within tasks. The system will automatically present key concepts in generated content in a graphical format, reducing users’ cognitive load through a visual hierarchy. Graphical representation will improve information readability, enhance memory retention, and enable users to more efficiently process complex tasks.

\paragraph{\textbf{DG4. Maintain hierarchical structure and enable logical classification and association of information.}}
Users need to maintain logical consistency and coherence in complex tasks, yet current systems underperform in information organization and hierarchical management (\textbf{DC4}). To address this, the system will use a structured hierarchical display to classify and associate information according to logical relationships, ensuring that users can better manage and retain key information during tasks. This structured design helps users clearly grasp relationships between topics, improves information organization efficiency, and allows users to customize the classification and association of information to meet different task requirements.

\section{System Design}

Based on the design goals outlined in Section 4, we developed a system that implements non-linear interaction and deep exploration while generating structured and visualized outcomes. This system is based on the "nodes + canvas" model, presented through a mindmap format, allowing users to freely explore and accomplish complex tasks.

\subsection{Node + Canvas System Design}

To enhance users' operational flexibility and freedom in exploring complex tasks, we designed a node-based non-linear interaction system with temporal and spatial flexibility, leveraging the capabilities of the ChatGPT-4.0 model to provide dynamic and contextually relevant responses~\cite{openai2023gpt}.

\subsubsection{Node-Based Non-linear Interaction}

To fulfill \textbf{DG1} (Provide non-linear interaction to enhance users’ freedom of exploration), the system employs a "nodes + canvas" design that breaks down long text paragraphs into independent nodes. Users can freely select, drag, and manipulate these nodes, removing the constraints of traditional linear interaction. This non-linear interaction enables users to jump between nodes and retrieve content flexibly, without following a fixed operational sequence. This greatly enhances operational flexibility, making it particularly suited for executing complex tasks.

\subsubsection{Temporal and Spatial Flexibility}

Addressing \textbf{DG1} further, the system provides flexibility in both time and space exploration. Using vector graphics rendering, it ensures node clarity even with large-scale information displays. Users can revisit previous operations or switch to different subtopics at any time, without being restricted by a linear task flow. The system supports zooming and dragging, enabling users to adjust their view and layout freely. This improves their ability to control tasks globally, especially in scenarios that require frequent shifts in focus.

\subsection{Node-Based Deep Exploration}

To support in-depth exploration, we provide AI-powered features and precise contextual generation that offer detailed, relevant information based on user needs.

\subsubsection{Deep Exploration Features}

In response to \textbf{DG2} (Provide deep information feedback to support in-depth exploration), the system offers three AI-powered features for each node—Explanation, Examples, and Exploration—allowing users to delve deeper into relevant information based on their needs. These features provide multi-dimensional support for task execution:

\begin{itemize} \item \textbf{Explanation}: Provides detailed concept explanations, enhancing users' understanding of node content. \item \textbf{Examples}: Displays three specific cases related to the node, using practical examples to improve comprehension. \item \textbf{Exploration}: Generates new responses based on user queries, expanding the user's exploration freedom. \end{itemize}

\subsubsection{Precise Contextual Generation}

To meet the expectations for deeper answers and exploration highlighted in \textbf{DG2}, the system uses predefined prompts tailored to different tasks to prevent overly broad or irrelevant answers, ensuring response precision. When users explore a node, the system inputs the parent node’s information as context, ensuring that the generated content is closely related to the background. Through iterative optimization, the system continually improves the accuracy of deep exploration and the efficiency of task execution.

\subsection{Information Visualization Design}

To simplify understanding of complex tasks, we employ visual design elements like color coding and optimized spatial layout to enhance information clarity and reduce cognitive load.

\subsubsection{Color Coding}

In line with \textbf{DG3} (Simplify understanding of complex tasks through graphical representation), the system uses color coding to mark nodes within the same subtopic, enhancing users' understanding of information hierarchy. Consistent visual cues enable users to quickly locate and distinguish related content, reducing cognitive load. This design is particularly effective for supporting users in handling multi-layered complex tasks, helping them easily identify and understand the structure of categorized information.

\subsubsection{Optimized Spatial Layout}

To further address \textbf{DG3}, the system optimizes the spatial layout to better present the logical relationships between nodes. By ensuring appropriate distance between nodes and their context, it avoids information overload or unclear associations. This spacing helps users visually grasp the hierarchical relationships between nodes. Compared to pure text-based paragraphs, the visualized node layout makes the information structure more logical, facilitating the handling of complex tasks.

\subsection{Hierarchical MindMap Structure}

To maintain hierarchical structure and logical consistency, we designed a multi-layered mindmap with parent-child associations and redundancy detection.

\subsubsection{Layered Information Presentation}

To satisfy \textbf{DG4} (Maintain hierarchical structure and enable logical classification and association of information), the system features a four-layer mindmap structure for initial generation. The first three layers use brief phrases to summarize core content, allowing users to quickly grasp the overall structure. The fourth layer provides detailed descriptions in complete sentences. This hierarchical presentation helps users swiftly identify key information and delve deeper into details when necessary, offering a clear framework for gradually refining complex tasks.

\subsubsection{Parent-Child Node Association and Redundancy Detection}

Addressing the need outlined in \textbf{DG4}, when generating child nodes via AI-powered features, the system ensures they are logically consistent with their parent nodes. Additionally, the system includes a redundancy detection feature to prevent the generation of duplicate content, ensuring that each node contains unique and valuable information. This strengthens users’ reliability in the system's outputs and improves task completion and logical rigor.

\section{Case Study}
One day, Anna, a master's student in statistics, came across the term "Surrealism" while reading. Curious about what it meant, she decided to explore this concept using the Mindalogue system. This exploration would guide her through the core ideas and influences of Surrealism while allowing her to experience the system's powerful features and flexible interactions.

\begin{figure*}[h]
  \centering
  \includegraphics[width=0.8\linewidth]{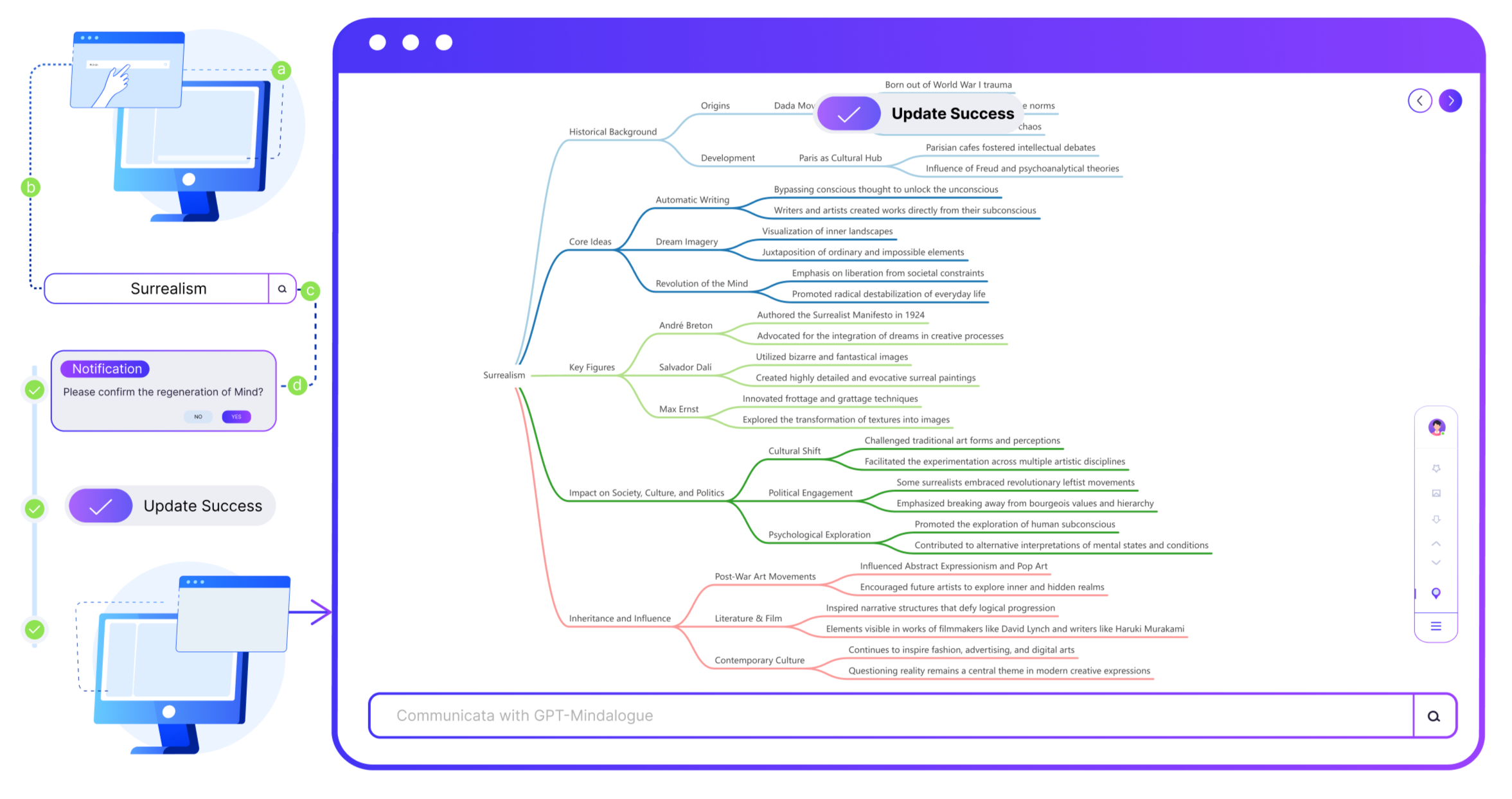}
    \vspace{-10pt}
  \caption{The user interaction in Mindalogue and the four levels of structured content that the system generates.}
  \label{Mindalogue System}
  \Description{This diagram illustrates the user operation process and the system page display. On the left, the user's operation process is shown in four steps: entering the keyword "Surrealism," system confirmation prompt, user confirmation, and successful update. On the right, the system's page display presents the final result as a mindmap.}
\end{figure*}

\subsection{Initial Exploration}

Anna entered "Surrealism" into the query box, and the system instantly generated a four-level mindmap (Figure ~\ref{Mindalogue System}). Each theme and sub-theme was clearly displayed with distinct colors and layers, helping Anna quickly grasp the structure of the concept.

Each node represented a key concept or sub-theme, ranging from the origins of Surrealism to its core ideas, major artists, and its impact on literature and culture. The color-coding made it easy for her to identify relationships between different levels, and the well-organized layout gave her a sense of order without feeling overwhelmed.

\subsection{In-Depth Exploration}

\subsubsection{Exploring with AI Functions}
\begin{figure*}[h]
  \centering
  \includegraphics[width=0.8\linewidth]{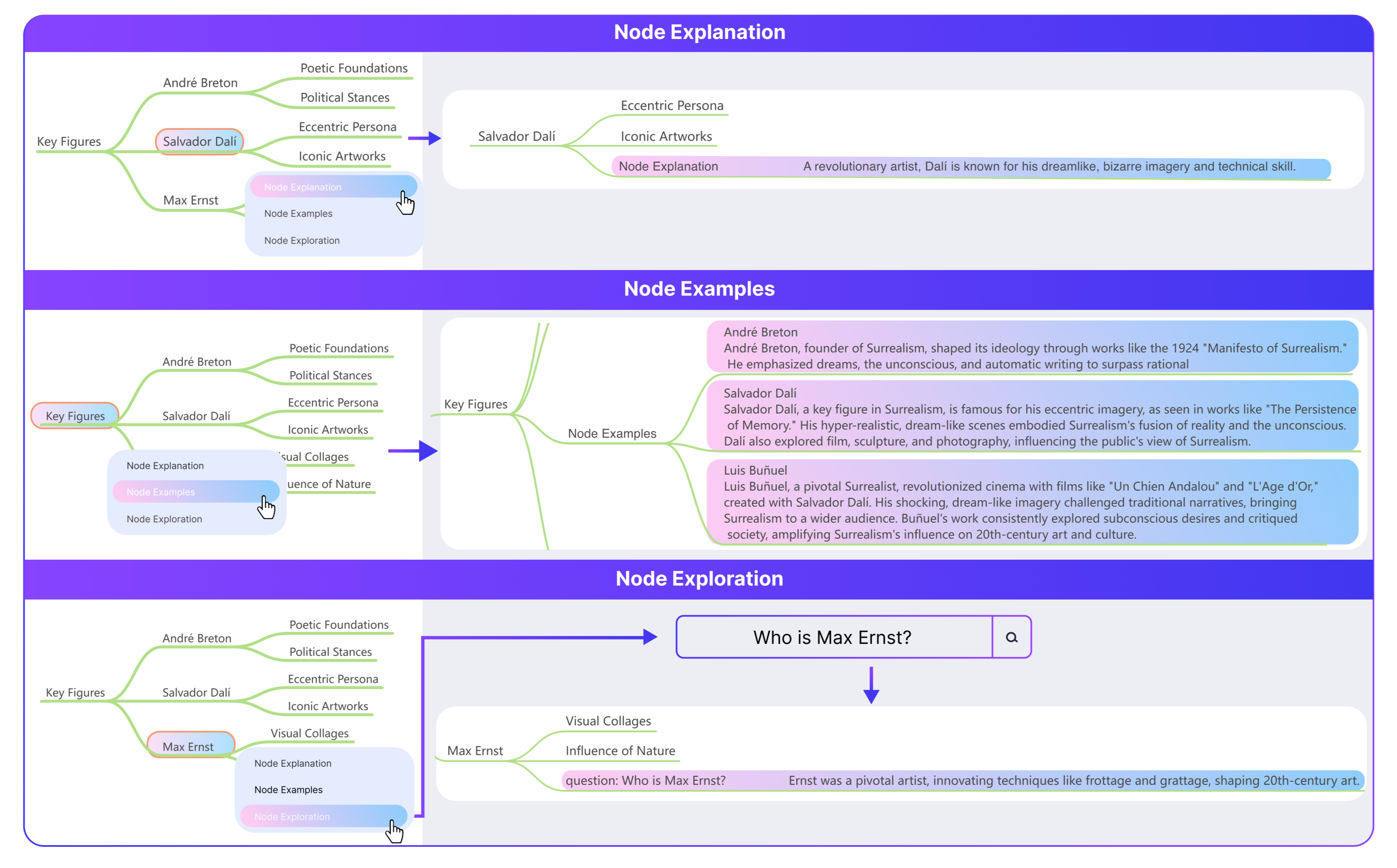}
  \vspace{-10pt}
  \caption{The three core functions of the Mindalogue system, including Node Explanation, Node Examples, and Node Exploration. Through these features, users are able to gain insight into the context and details of concepts, supporting multiple levels of content interaction and exploration.}
  \label{ai functions}
  \Description{This diagram illustrates the system's functional operations and result presentation. It highlights the three main functions: Node Explanation, Node Examples, and Node Exploration. On the left, the user's operation process is shown as they select each function in sequence. On the right, the corresponding result for each operation is displayed.}
\end{figure*}

Intrigued, Anna started with the "Salvador Dali" node. Right-clicking on the node revealed three AI exploration options (Figure ~\ref{ai functions}). She chose the explain function, which provided detailed information about Dali's background and artistic style.

Next, Anna wanted to explore the specific styles and works of Surrealist artists. She clicked on examples, and the system quickly presented three case studies, including André Breton's Surrealist Manifesto and Dali’s iconic works. These examples deepened her understanding of the concept.

Still curious, Anna noticed another artist, Max Ernst, and used the explore function to ask, "Who is Max Ernst?" The system generated a detailed response, giving her further insights into the connections and characteristics of Surrealist figures.

\subsubsection{Custom Exploration}

\begin{figure*}[t]
  \centering
  \includegraphics[width=0.8\linewidth]{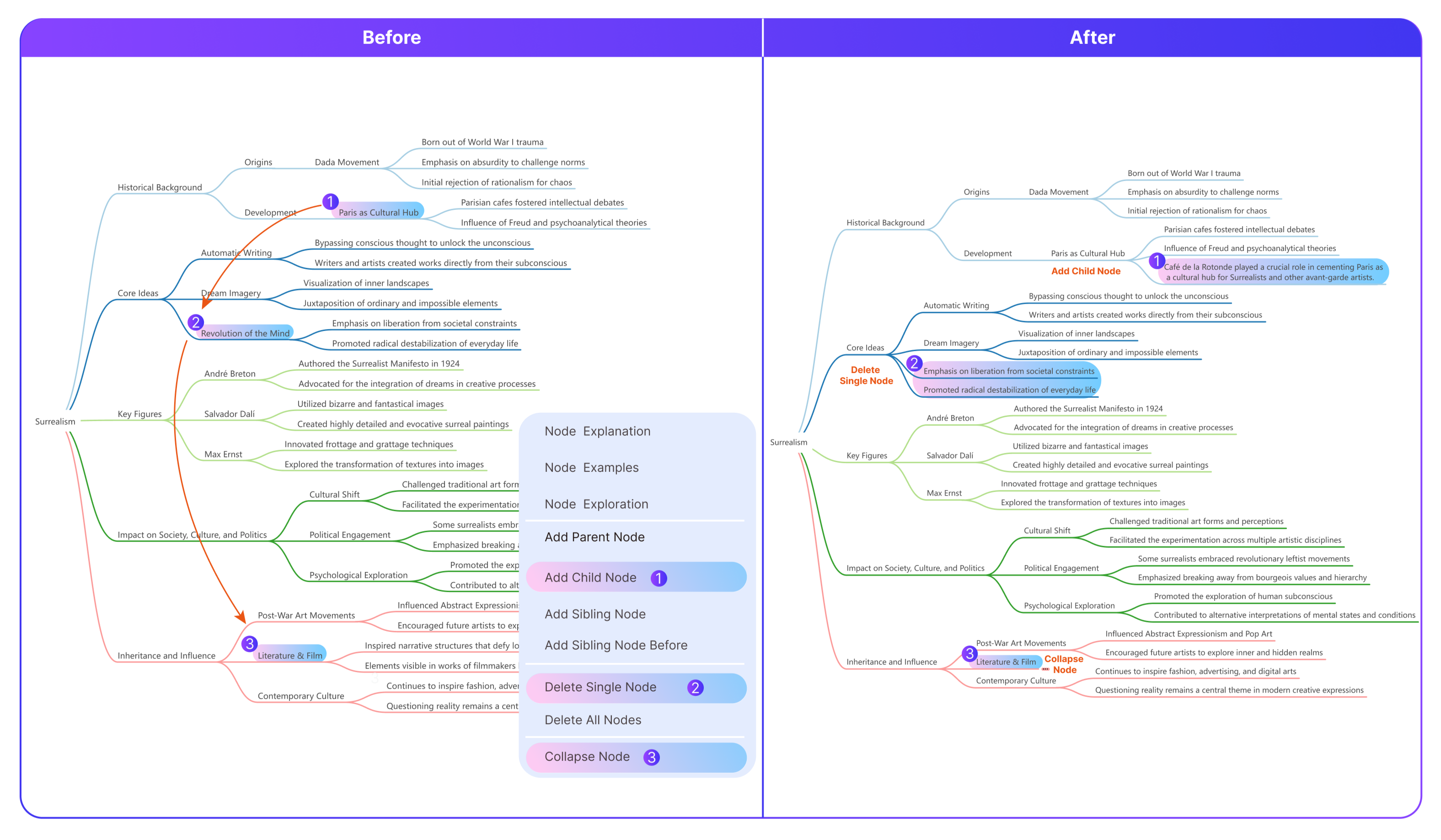}
    \vspace{-10pt}
      \caption{Mindalogue system's custom exploration feature allows users to flexibly explore and adjust the structure of the content by adding, deleting, and collapsing nodes and other operations for non-linear interaction.}
  \label{custom functions}
  \Description{This comparison chart illustrates the effects of three randomly selected customization operations. On the left is the mindmap before the operations, and on the right are the changes after the operations. The user randomly selected "add child node," "delete single node," and "collapse node," with the right side showing the resulting modifications to the mindmap. The flexible navigation on the left highlights the advantages of non-linear interaction, allowing users to move seamlessly between different nodes without being constrained by a linear process.}
\end{figure*}

In addition to the provided information, Anna wanted to incorporate her own insights. Using the customization feature, she added a new node to reflect her understanding. She also edited and removed redundant or irrelevant nodes to better fit her needs (Figure ~\ref{custom functions}). The system allowed her to easily modify the mindmap structure, and with the undo and redo buttons, she could quickly reverse or advance her actions as needed.

\subsection{Interaction Experience}

\subsubsection{Non-linear Interaction}

As Anna explored further, she realized she wasn’t bound by any fixed sequence. She could freely jump between nodes. After studying "Paris as a cultural hub," she moved directly to "Revolution of the mind," and later explored "Literature \& Film." This non-linear interaction gave her the flexibility to explore without the constraints of a traditional linear structure.

\subsubsection{Collapsing and Expanding the Structure}

As the mindmap became more complex, Anna used the system’s collapse function to temporarily hide certain nodes, simplifying her view and reducing cognitive load. When she needed more details, she used the expand function to reopen those nodes, ensuring quick access to additional information when necessary.

\section{Evaluation Study}

The primary objective of this study is to compare the non-linear interaction system, Mindalogue, with the linear interaction system, both of which are built on the GPT-4o model. Specifically, we aim to investigate whether non-linear interaction can significantly enhance user freedom in task execution, improve operational flexibility, reduce cognitive load, and expand the exploration space. By analyzing user behavior and feedback during complex task exploration, we seek to validate whether graphical structures effectively support the processing and exploration of complex information.

\subsection{Participants}

We recruited 16 participants to evaluate two systems for completing practical complex tasks. There were 6 males and 10 females, aged 21 to 30 (mean = 24.06, SD = 2.52). Regarding LLMs usage frequency, 3 were monthly users, 5 were weekly users, and 8 were daily users. Detailed participant information is provided in Table~\ref{tab:participant}. Each participant spent approximately 120 minutes total in our study, and received euqal compensation for their time.

\begin{table*}[t]
  \caption{Participant Background}
  \vspace{-10pt}
  \label{tab:participant}
  \resizebox{0.7\linewidth}{!}{
  \begin{tabular}{clllcc}
    \toprule
    Participant & Gender & Age & Academic Background & Education Level & LLMs Usage Frequency \\
    \midrule
    P1  & Female & 21 & Environmental Science & Bachelor & Daily \\
    P2  & Female & 26 & Law \& Social Work & Master & Weekly \\
    P3  & Female & 24 & Science \& Finance & Master & Weekly \\
    P4  & Male   & 30 & Mechanical Engineering & Master & Weekly \\
    P5  & Female & 25 & Sociology \& User Research & Master & Daily \\
    P6  & Female & 24 & Film and Multimedia & Master & Daily \\
    P7  & Male   & 26 & Chemistry \& Design & PhD & Daily \\
    P8  & Female & 28 & Telecommunications \& Agronomy & PhD & Weekly \\
    P9  & Male   & 22 & Medicine & Bachelor & Monthly \\
    P10 & Female & 22 & Design & Bachelor & Daily \\
    P11 & Female & 22 & Industrial Design & Master & Daily \\
    P12 & Male   & 22 & Cultural Management & Bachelor & Weekly \\
    P13 & Male   & 22 & Industrial Design & Master & Monthly \\
    P14 & Male   & 22 & Artificial Intelligence & Master & Daily \\
    P15 & Female & 24 & Musicology \& Art Theory & Master & Monthly \\
    P16 & Female & 25 & Science \& International Trade  & Master & Daily \\
    \bottomrule
  \end{tabular}
  }
  \Description{The table presents the background information of the participants, including gender, age, academic background, education level, and the frequency of LLMs usage. A total of 16 participants were involved, with a gender distribution of 10 females and 6 males, and an age range from 21 to 30 years. The academic backgrounds of the participants span multiple fields, such as Environmental Science, Law and Social Work, Mechanical Engineering, Sociology, and User Research, indicating a diverse sample. In terms of education level, 10 participants hold master's degrees, 4 have bachelor's degrees, and 2 are currently pursuing doctoral studies, reflecting a generally high level of education among the participants. Regarding LLMs usage frequency, 8 participants use LLMss daily, 5 use them weekly, and 3 use them monthly. This table provides foundational data for the study, contributing to an understanding of LLMs usage across different academic backgrounds and its potential implications.}
\end{table*}

\subsection{Setup}

\subsubsection{Systems}

The experiment compared two systems. Mindalogue (System 1) was deployed on a cloud server and used a non-linear interaction for information organization and exploration. Mindmap Master (System 2) used a linear interaction for generating responses based on the ChatGPT-4o model. Both systems were initialized with identical settings to ensure the accuracy of the comparison (see Appendix's A.Prompts). The key feature of System 1 lies in its ability to allow users to freely explore and reorganize information through a graphical structure, while System 2 primarily relies on sequential text-based interaction. Both systems' responses were generated by the ChatGPT-4o model. This comparison directly reflects the systems' capabilities in supporting users through complex tasks.

\subsubsection{Tasks}

We designed four tasks: two tasks on unfamiliar topics (Knowledge Exploration and Brainstorming) and two tasks on familiar topics (Product Analysis and City Description)(Table~\ref{tab:task_settings}). Each participant completed all four tasks. For the unfamiliar tasks, the Knowledge Exploration topics were Machine Learning, Dadaism, Relativity, and the Industrial Revolution. These tasks assessed whether the systems could provide sufficient depth of information to support effective user learning. The Brainstorming tasks covered Future Fashion Design, Commercial Dream Experiences, Digital Food Design, and Event Planning Without Digital Products, focusing on how well the systems could facilitate creative thinking. For the familiar tasks, participants recommended a commonly used mobile app, testing the system's logical and accurate analyses. In the City Description task, participants recommended a city, evaluating the system's ability to offer detailed and comprehensive information.

To ensure balanced sample sizes across all topics, participants were assigned different topics in sequence and completed two tasks on each system for comparison (Supplementary Material). A total of 16 participants completed the study, with 16 samples for each task and 8 samples per task for each system. To maintain unfamiliarity with the topics in Task Category 1 - Knowledge Exploration, topics were assigned based on the participants' backgrounds: participants with a STEM background were given humanities-related topics, while those with a humanities background were assigned STEM-related topics. This cross-assignment ensured that participants approached each task with a similar level of unfamiliarity.

\begin{table*}[t]
  \caption{Task Settings}
  \vspace{-10pt}
  \label{tab:task_settings}
  \begin{tabular}{lll}
    \toprule
    Task Category & Topic & Familiarity \\
    \midrule
    C1: Knowledge Exploration & Machine learning, Dadaism, Relativity, Industrial Revolution & Unfamiliar \\ 
    C2: Brainstorming & Future fashion design, Commercial dream experience, & {Unfamiliar} \\
                           & Digital food design, Event planning without digital products &  \\ 
    C3: Product Analysis & Recommend a commonly used mobile app & Familiar \\ 
    C4: City Description & Recommend a city (e.g., a favorite city, hometown) & Familiar \\
    \bottomrule
  \end{tabular}
  \Description{The table provides a detailed outline of the specific settings for various tasks, encompassing four categories: knowledge exploration, brainstorming, product analysis, and city description. In the knowledge exploration (C1) and brainstorming (C2) tasks, participants are required to engage with unfamiliar themes. The knowledge exploration task includes a diverse array of topics such as machine learning, dadaism, relativity, and the industrial revolution. In contrast, the brainstorming task covers innovative subjects like future fashion design, commercial dream experience, digital food design, and event planning without digital products. Conversely, the product analysis (C3) and city description (C4) tasks focus on themes familiar to the participants. The product analysis task entails recommending a commonly used mobile app, while the city description task requires participants to recommend a city with which they are familiar.}
\end{table*}

\subsubsection{Video Recording}

All remote testing sessions were conducted using Tencent Meeting, with screen recording initiated at the start of each session to capture participants' screen activity and interactions. These recordings were transcribed, and detailed analyses of user behavior, task completion paths, and interaction patterns were conducted. Finally, we conducted thematic analysis on the semi-structured interview content to gain qualitative insights.

\subsection{Procedure}
\begin{figure*}[t]
  \centering
  \includegraphics[width=0.8\linewidth]{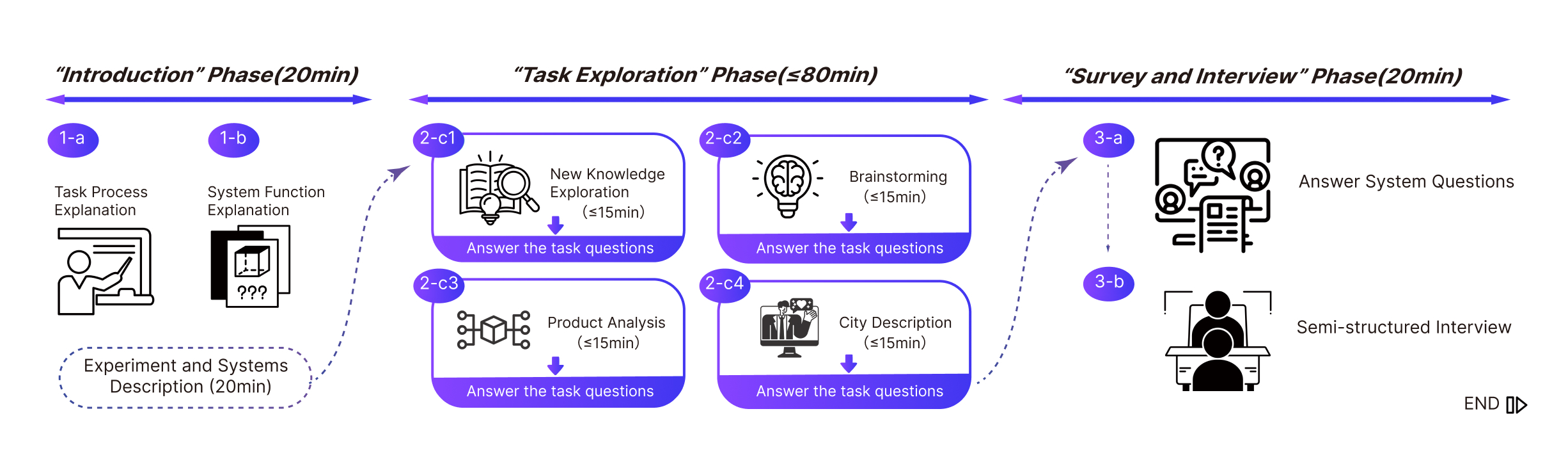}
  \vspace{-10pt}
  \caption{Evaluation Study Procedure with 3 Phases.}
  \label{procedure}
  \Description{The experimental flow chart highlights three main stages: "Introduction Phase" (<20 min), "Task Exploration Phase" (<80 min), and "Survey and Interview Phase" (<20 min). In the "Introduction Phase," the experimenter explains the task process and system functions to participants. The "Task Exploration Phase" follows, consisting of four tasks: New Knowledge Exploration, Brainstorming, Product Analysis, and City Description, each taking up to 15 minutes. Participants answer task-related questions after each exploration. The final stage, "Survey and Interview Phase," involves completing system-related questions and a semi-structured interview.}
\end{figure*}

The study session had 3 phases: introduction, task exploration, and survey interview, as shown in Figure~\ref{procedure}.

\subsubsection{Introduction(20 minutes)}
The experimenter briefly introduced the study's objectives. After participants signed the informed consent form, they were given an overview of the experiment process and the four tasks to be completed. The experimenter then introduced the usage of both systems and demonstrated a non-test topic (e.g., Surrealism) to guide users through the interfaces and features of Mindalogue(S1) and Mindmap Master(S2). Participants were then given 10 minutes to explore the systems using the demonstration topic, allowing them to familiarize themselves with the interface and functionality for an optimal experience.

\subsubsection{Task Exploration(80 minutes)}

Guided by the experimenter, users completed the assigned tasks on both the S1 and S2 systems. Specifically, in Tasks 1 and 2, users acted as learners/explorers, using the systems to support their learning of new knowledge and the design of solutions. In Tasks 3 and 4, users acted as recommenders or describers, selecting a mobile app and a city, respectively, and using the systems to structure and enrich their presentations.
To control the total test duration, we set a maximum completion time of 15 minutes for each task. After completing each task, users had 1-5 minutes to answer a task-specific quantitative question. These task-related questionnaires included four questions, one of which was from the CSI scale, aimed at assessing how well the different systems supported the completion of complex tasks.

\subsubsection{Survey and Interview(20 minutes)}

After completing the tasks, participants filled out a comparative questionnaire consisting of 9 questions across five dimensions: convenience, usability, controllability, cognitive load, and result reliability. Each dimension had 1-3 questions, including 5 items from the SUS and NASA-TLX scales. Finally, we conducted an semi-structured interview, asking participants to reflect on seven dimensions: structure, flexibility, AI reliability, interaction and interface design, cognitive load, system recommendations, and other perspectives. The goal was to compare the strengths and weaknesses of the two systems, understand their limitations, and explore potential applications. Detailed questions are provided in Supplementary Material.

\subsection{Method}

We compared the interaction data between the two systems through descriptive statistical analysis, focusing on task duration and the number of actions. Task duration was measured from the beginning to the end of each task, while the number of actions was defined as all participant-initiated interactions within the system. In System 1, total actions included mind map generation, AI exploration actions (explanation, examples, exploration), and participant custom actions (add, delete, edit, move). In System 2, the number of actions was calculated based on participant-initiated dialogues. For task-related questions, we used means and standard deviations to perform a comparative analysis of system support. To ensure the internal consistency of system comparison questions, we used Cronbach’s alpha, and paired t-tests were conducted to calculate the significance (p-value) of differences between systems. For the semi-structured interview data, we transcribed the audio recordings and conducted qualitative analysis to extract key insights that support the conclusions drawn from the quantitative data.

\section{Results}

\subsection{Quantitative results}

\subsubsection{Mindalogue Provides Users with Greater Exploration Space and Engagement}

By comparing the task completion time and number of actions in S1 (Mindalogue) and S2 (traditional system), we found that users spent more time and performed more actions in S1. For the four task categories, the average time spent in S1 was (M = 10.16, SD = 3.20), while in S2 it was (M = 7.78, SD = 2.50). In terms of total actions, S1 recorded (M = 11.19, SD = 4.90), and S2 recorded (M = 6.81, SD = 3.04). These results indicate that the Mindalogue system offers users a larger exploration space, allowing them to explore more freely and demonstrate greater curiosity in engaging with the content.

\subsubsection{Unfamiliar Tasks Require More Exploration}

When comparing familiar and unfamiliar topics, we found that users completed tasks more quickly on familiar topics, while they took more time and performed more actions on unfamiliar ones. This suggests that users need additional time to understand and adapt to unfamiliar tasks. For unfamiliar topics (Tasks 1 and 2), the average exploration time was S1 (M = 12.13, SD = 2.83) and S2 (M = 8.94, SD = 2.93). The total number of actions was S1 (M = 12.44, SD = 5.63) and S2 (M = 7.31, SD = 3.10). For familiar topics (Tasks 3 and 4), the average exploration time was S1 (M = 8.19, SD = 3.57) and S2 (M = 6.63, SD = 2.06), with total actions recorded as S1 (M = 9.94, SD = 4.18) and S2 (M = 6.31, SD = 2.98). These findings suggest that users are more efficient with familiar topics, while unfamiliar topics require more exploration and actions.

\subsubsection{Mindalogue's Superior Performance in Complex Task Support}

\begin{figure}[t]
  \centering
  \includegraphics[width=\linewidth]{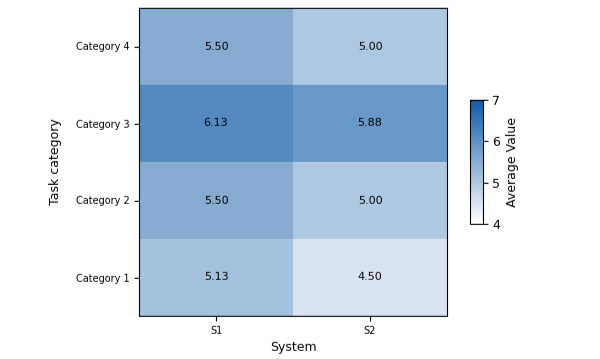}
  \caption{The heatmap compares the average user ratings (1-7 scale, higher is better) for Systems 1 (S1) and 2 (S2) across four task categories. S1 represents System 1 and S2 represents System 2. The ratings reflect how well each system supports the given task, with values indicating the mean scores for each category.}
  \label{Tasks Comparison}
  \Description{The figure illustrates the average values of systems S1 and S2 across different task categories. The vertical axis represents the task categories (from Category 1 to Category 4), while the horizontal axis denotes the systems (S1 and S2). The heatmap displays the average values of each task category within the different systems using a color gradient, with a numerical range from 4 to 7, transitioning from light to dark shades. The results indicate that the average values for system S1 are consistently higher than those for system S2 across all task categories, with the most significant difference observed in Category 3 (S1: 6.13, S2: 5.88).}
\end{figure}

Based on the comparison of the four task categories in terms of learning support, creative support, logical classification, and descriptiveness, the average scores for S1 consistently outperform those for S2 (Figure~\ref{Tasks Comparison}). This indicates that Mindalogue's non-linear interaction model is more advantageous in handling complex tasks compared to the traditional linear interaction of LLMs.

Looking at the data across the four task, users rated both systems highest when completing Task 3 (product decomposition). S1 demonstrated stronger logical structuring abilities (M=6.13, SD=0.64), while S2's performance in this area was slightly lower (M=5.88, SD=0.83). For Task 2 (brainstorming) and Task 4 (describing a city), both systems received similar scores. For Task 2, S1 (M=5.5, SD=1.41) and S2 (M=5, SD=1.51) showed comparable support, while in Task 4, S1 (M=5.5, SD=1.07) and S2 (M=5, SD=1.31) were equally matched.
The lowest system support across all tasks was observed in Task 1 (knowledge exploration). S1 scored an average of (M=5.13, SD=1.13), while S2 received a lower score of (M=4.5, SD=1.2). This suggests that both systems were less effective in supporting users with knowledge acquisition tasks.

\subsubsection{Mindalogue Excels in Flexibility, Reliability, Controllability, and Trustworthiness}

To assess the reliability of the scale, we calculated Cronbach's Alpha. The coefficient for S1 was 0.74, and for S2 it was 0.84, both exceeding the standard threshold of 0.7~\cite{tavakol2011making}, the data from both systems exhibit good internal consistency.

\begin{figure}[t]
  \centering
  \includegraphics[width=\linewidth]{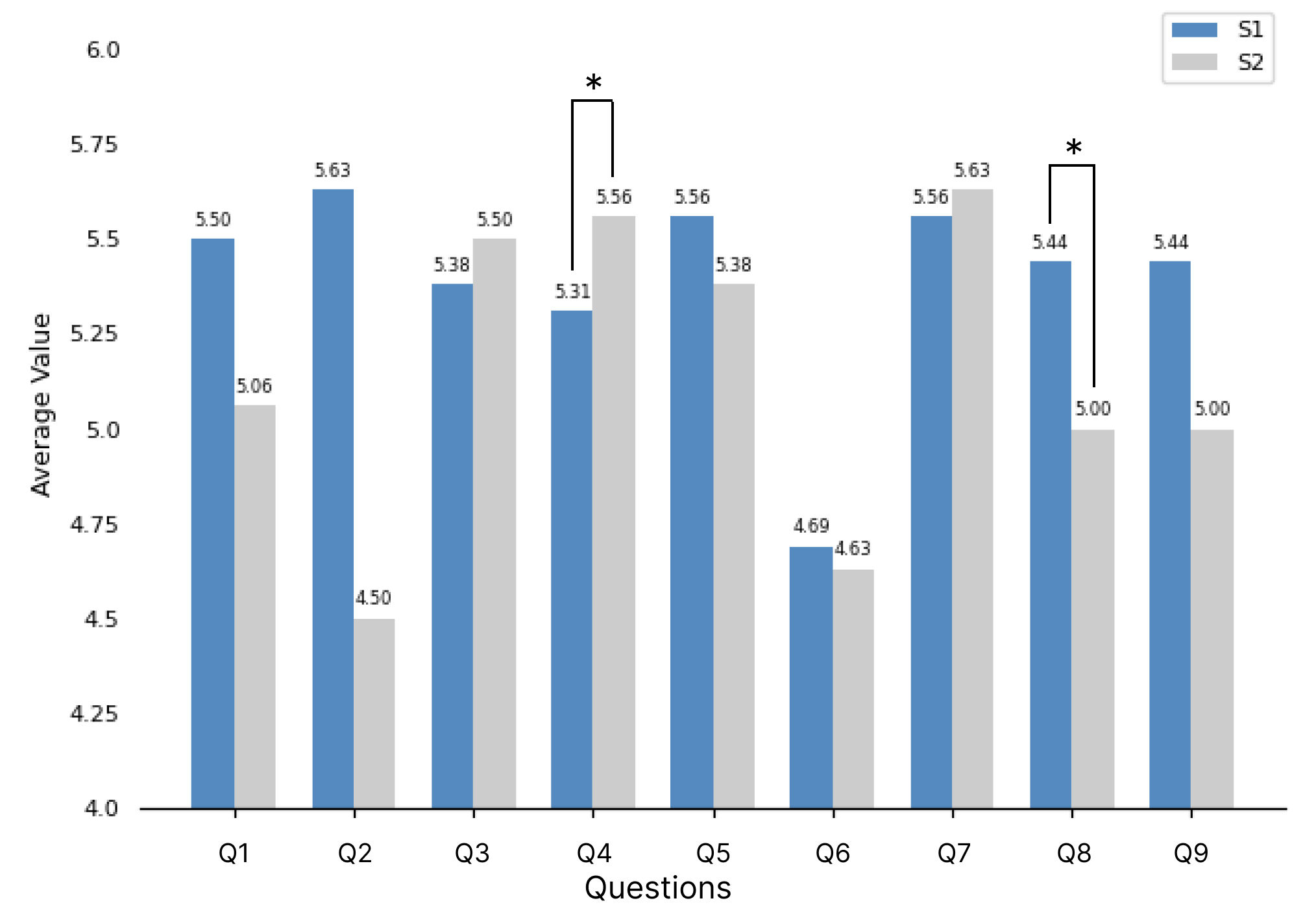}
  \vspace{-10pt}
  \caption{The bar chart presents average user ratings for 9 questions across two systems, S1 and S2. The vertical axis shows the average value, with values set between 4 and 6 to highlight the differences between systems. Higher values indicate better performance. S1 performs better than S2 on  Q1, Q2, Q5, Q6, Q8, and Q9, with a statistically significant difference in favor of S1 on question 8 (p < 0.05). However, S1 performs slightly worse than S2 on Q3, Q4, and Q7, with S2 significantly outperforming S1 on question Q4 (p < 0.05).}
  \label{System Comparison}
  \Description{The figure presents a comparative analysis of the average values for two systems (S1 and S2) across nine categories of questions (Q1 to Q9). The results indicate that S1 outperforms S2 in the majority of categories, particularly in Q1, Q2, Q8, and Q9, where the average values for S1 are 5.50, 5.63, 5.44, and 5.44, respectively, compared to S2's corresponding values of 5.06, 4.50, 5.00, and 5.00. Although S1's average values are lower than those of S2 in the categories Q3, Q4, and Q7, the differences between the two systems are relatively small.}
\end{figure}

\paragraph{Mindalogue Excels in Flexibility and User Satisfaction in Most Dimensions}
We used paired-samples t-tests~\cite{manfei2017differences}to compare the mean scores and significance differences between the two systems across 9 questions (Figure~\ref{System Comparison}). Among the 9 sub-dimensions covered by the questions, 6 had higher mean scores for S1 compared to S2 (greater operational flexibility, higher content freedom, system functionality meeting expectations, lower psychological difficulty during task completion, higher satisfaction with task completion, and greater reliability in the final outcome). Meanwhile, S2 outperformed S1 in 3 sub-dimensions (greater confidence in using the product, lower learning effort required before using the system, and less perceived time pressure during task completion).

\paragraph{Mindalogue Enhances User Confidence and Satisfaction in Complex Task Exploration}

Specifically, question 4 had a p-value of 0.008, indicating that S2 requires less learning effort. This aligns with the fact that LLMs, being dialogue-based and having a simpler linear interaction format, are easier to learn. Additionally, all 16 participants had prior experience with LLM products, which may explain why S2 was rated lower in terms of learning cost. Question 8 also showed a significant difference (p = 0.033), with users expressing significantly higher satisfaction when completing tasks in S1. Mindalogue appears to give users greater confidence and satisfaction in their task performance.

\paragraph{Mindalogue Outperforms in Convenience, Reliability, Cognitive Load and Controllability}

\begin{figure}[h]
  \centering
  \includegraphics[width=\linewidth]{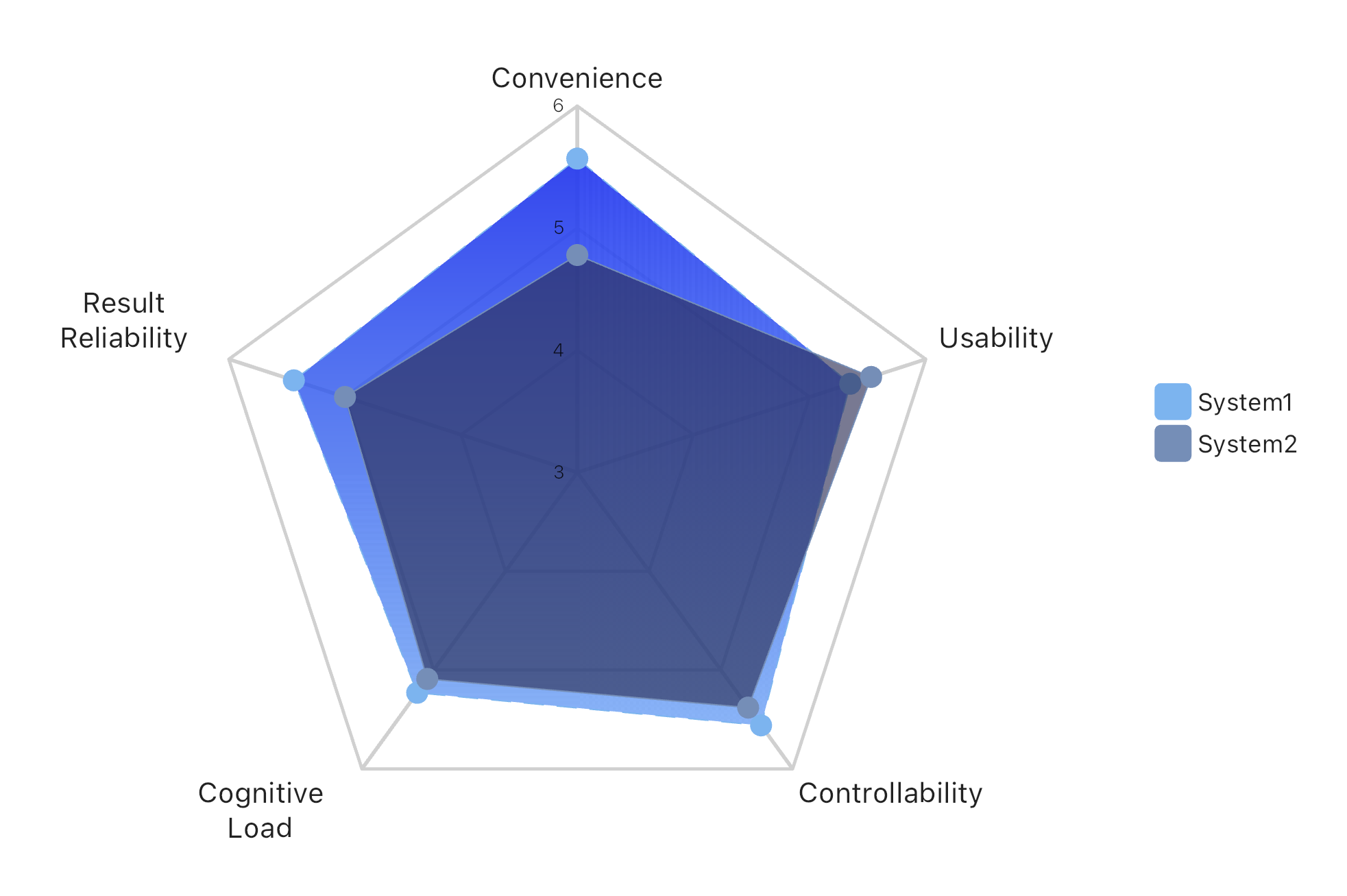}
  \caption{The radar chart compares user ratings of System 1 and System 2 across five dimensions: Convenience, Usability, Controllability, Cognitive Load, and Result Reliability, with values set between 3 and 6 to highlight the differences between systems. Higher values indicate better performance, with System 1 outperforming System 2 in 4 dimensions, except for Usability.}
  \label{Radar Chart}
  \Description{The figure illustrates the performance of two systems (S1 and S2) across five evaluation dimensions: Convenience, Usability, Result Reliability, Controllability, and Cognitive Load. The results presented in the radar chart indicate that S1 outperforms S2 in all evaluation dimensions except for Usability. Notably, S1 demonstrates a significant advantage in both Convenience and Result Reliability.}
\end{figure}

By calculating and processing the average values of each sub-dimension, we generated a radar chart comparing the two systems across the five key dimensions: convenience, usability, sense of control, cognitive load, and result reliability (Figure\ref{Radar Chart} ). The comparison reveals that Mindalogue (S1) exhibits a more balanced and stable performance across the five dimensions. S1 outperforms S2 in four dimensions: convenience, sense of control, cognitive load, and result reliability. Notably, the largest difference is observed in the convenience dimension, indicating that users experience greater flexibility and freedom when using Mindalogue. However, in the usability dimension, S1's average score is slightly lower than S2, suggesting that the non-linear interaction of Mindalogue may involve a higher learning curve compared to the traditional linear interaction of S2. Despite this, once users become familiar with the system, S1 offers a larger exploration space.

\subsection{Quanlitative results}

\subsubsection{Non-linear Interaction Supports Flexible Operation and Free Exploration}

S1 allows users to add and edit nodes, freely search for information, and choose paths, supporting divergent thinking and deep exploration, thereby improving task efficiency. For example, P4 stated that S1 enables \textit{"arbitrary exploration,"} with generated answers displayed \textit{"visually"} on the graph (P4, P9). P9 mentioned that content editing is \textit{"very flexible,"} enhancing efficiency (P4, P9). Similarly, P3 observed that S1's node operations and interactive questioning are highly flexible (P3). In contrast, S2 lacks flexibility and freedom, limiting users' divergent thinking. P2 felt that S2 is \textit{"limited"} in free questioning and customization, especially inconvenient in content editing (P2). P6 pointed out that S2's linear limitations are \textit{"quite restrictive"} in complex tasks (P2, P6).

\subsubsection{Enhancing Understanding of the Overall Task}

Users found that S1's mindmap format enhances their understanding of the overall task, providing clear logical structure and hierarchy. For example, P1 stated that node explanations and case functions \textit{"helped the most"} in quickly grasping the entire task (P1, P3, P4). P3 mentioned that the graphical mindmap aids in \textit{"quickly understanding the whole task"} (P1, P3, P4). P4 believed that tree diagrams allow for \textit{"clearer"} comprehension, improving efficiency (P1, P3, P4). Conversely, S2's linear Q\&A mode lacks a clear logical structure, making it difficult for users to quickly understand the overall task (P3). P3 pointed out that S2's vertical browsing is \textit{"not very convenient"} (P3). P6 felt that S2's linear structure is \textit{"limited"} in its presentation (P3, P6).

\subsubsection{Structured Results Increase Reliability}

Through structured presentation and specific case explanations, S1 enhances users' reliability in the system-generated content, especially on unfamiliar topics. P3 stated that the example explanation function boosts reliability in AI because \textit{"specific case explanations"} make the content more credible (P3, P4). P4 believed that S1's case analysis is \textit{"more convincing"} (P3, P4). In contrast, S2's responses may have accuracy and reliability issues, reducing users' reliability. P2 pointed out that S2's answers are \textit{"too repetitive, affecting reliability"} (P2, P9). P9 felt that S2's responses are \textit{"too unreliable,"} more like \textit{"chatting with a search engine"} (P2, P9).

\subsubsection{Reducing Cognitive Load}

S1's classification and chart functions help users organize their thoughts, reduce cognitive load, and facilitate handling information-intensive tasks. P1 mentioned that the classification function \textit{"reduced cognitive load,"} hoping it could highlight key points more effectively (P1, P3). P3 stated that S1's chart function is \textit{"very good,"} helping users organize thoughts and display information clearly (P1, P3). In contrast, when dealing with complex information, S2 lacks a clear logical structure, increasing users' cognitive load. P3 pointed out that S2's content is sometimes \textit{"quite complex, with unclear relationships"} (P3, P9). P9 believed that S2 lacks detail, offering only \textit{"broad categories without subcategories,"} making understanding difficult (P3, P9).

\subsubsection{Suitable for Complex and Divergent Tasks}

S1 performs excellently in tasks requiring structured and divergent thinking, such as learning new knowledge, brainstorming, and product analysis, providing greater operational flexibility and freedom. P6 believed that S1 is very suitable for \textit{"previewing or reviewing unfamiliar knowledge points,"} able to generate a complete system framework (P6, P9). P9 stated that S1 is more helpful in \textit{"understanding things,"} especially when constructing an overall framework is needed (P6, P9). In contrast, S2 has limitations in handling complex tasks and cannot meet users' needs for in-depth exploration (P6, P16). P6 pointed out that S2 appears \textit{"quite limited"} in complex tasks, with content being too linear (P6, P16). P16 felt that S2's linear structure is \textit{"not flexible enough"} in divergent tasks and needs more features to support divergent thinking (P6, P16).

\subsubsection{Limitations and Suggestions}

\paragraph{Desire for Enhanced Editing and Marking Functions}
Users hope that S1 can improve its interface and interaction by adding more editing and marking functions. P1 wished for \textit{"marking functions,"} such as changing colors and bolding fonts (P1, P8). P8 mentioned hoping to \textit{"automatically generate sub-nodes"} and further enhance branch color distinctions to improve the user experience (P1, P8). P3 suggested improving undo and error-tolerance features to avoid \textit{"mistaken operations disrupting the mindmap"} (P3).

\paragraph{Content Repetition and Information Redundancy}
Some users pointed out that S1 may have issues with content repetition and information redundancy. P5 stated that during node expansion, \textit{"content repetition is quite high,"} affecting the depth of tasks (P5, P6). P6 believed that there is a need to improve content generation guidance to avoid randomness and redundancy in information (P5, P6).

\paragraph{Need for Additional Features}
Users also expect S1 to provide image displays and reference sources to enhance usability. P2 hoped to add a \textit{"generate images"} function (P2, P11, P14). P11 mentioned that examples would be better if they included \textit{"image displays"} (P2, P11, P14). P14 suggested that adding images when giving examples makes it \textit{"clearer and more understandable"} (P2, P11, P14).

\subsection{Summary}

Mindalogue provides a larger exploration space, and users are willing to perform more operations to explore content in depth. They experience greater flexibility, content freedom, and satisfaction in completing tasks. Mindalogue outperforms traditional linear systems in convenience, sense of control, reduced cognitive load, and result reliability. Qualitative feedback reinforces these findings, highlighting Mindalogue's strengths in supporting flexible operation, enhancing overall task comprehension, reducing cognitive load, and increasing reliability through structured results. Despite a slightly higher learning curve, users find that once they become familiar with the system, Mindalogue offers a more effective platform for complex task exploration and learning. Future efforts will focus on improving usability features—such as enhancing editing functions and reducing content redundancy—to further refine performance and user experience.
\section{Discussion}
Our study experimentally compared the performance of the non-linear interaction system, Mindalogue, with traditional linear interaction systems, showcasing the significant advantages of Mindalogue in task decomposition, freedom of information exploration, cognitive load reduction, and graphical representation~\cite{shneiderman1992tree,card1991information,kirsh2001context}. However, we also identified certain technical and interaction-related limitations. The following discussion will analyze the strengths, limitations, and potential future applications of the system based on experimental findings.

\subsection{System Advantages and Areas for Improvement}
\subsubsection{Enhancing Task Decomposition and Exploration Efficiency}
The "nodes + canvas" structure of Mindalogue provides a non-linear flow of information, enhancing users' ability to decompose and explore complex tasks. Experimental results indicate that participants demonstrated greater flexibility when using Mindalogue, enabling them to manage and handle multi-layered tasks simultaneously~\cite{bederson2000fisheye}. This design effectively reduced the cognitive load on users as they progressed through tasks, offering more possibilities for interdisciplinary exploration~\cite{kieras1985approach,ware2002cognitive,kirsh2001context}. In the future, intelligent guidance (e.g., dynamically suggesting the next step or interaction path based on user operations) could be introduced to help users decompose tasks more efficiently, reducing the cognitive load associated with complex interactions~\cite{shneiderman2003eyes}.

\subsubsection{Improving the Clarity of Information Structures}
Mindalogue’s graphical representation allows complex information to be displayed in hierarchical structures, helping users understand the relationships between concepts more intuitively~\cite{card1991information}. Qualitative feedback from participants shows that the graphical format improved their grasp of information structures, especially when dealing with tasks involving multiple dimensions and layers~\cite{clarinval2018evoq}. However, some users reported that the complexity of nodes increased operational burden in certain cases~\cite{ware2002cognitive}. In the future, progressive disclosure techniques could be implemented, allowing users to adjust the level of information displayed according to their needs, avoiding information overload and enhancing the overall user experience~\cite{bederson2000fisheye}.

\subsection{Challenges and Limitations}
\subsubsection{Issues with Annotation and Information Accuracy}
Despite the strong performance of Mindalogue in generating annotations, occasional errors or inconsistencies were observed in complex, multi-layered relationships~\cite{chang2002effect}. For instance, the system might incorrectly generate relationships between a node and a non-existent entity or misinterpret the actual association between two nodes, leading to graphical annotations pointing to inaccurate information~\cite{shneiderman1992tree}. These errors can confuse users during task operations, requiring additional time for verification and correction~\cite{clarinval2018evoq}. To address this issue, future development could integrate domain-specific knowledge bases to improve annotation accuracy, and an automated validation mechanism could be introduced to ensure that generated annotations are cross-referenced and verified against external data sources, reducing the likelihood of misleading information. Such a mechanism would help increase user reliability in system-generated content, particularly when handling complex and professional domains~\cite{hornbaek2011notion}.

\subsubsection{Complexity of Non-linear Interaction}
Although non-linear interaction provides users with greater flexibility and freedom, its high degree of freedom also poses challenges for some users, especially in information-dense tasks~\cite{hornbaek2011notion}. Some participants in the experiment reported that they easily lost direction when dealing with complex node networks, particularly when managing a large volume of information. To address this, more guided prompts and interaction path optimizations could be introduced in interface design. For example, intelligent task/function guidance could be provided, automatically recommending interaction paths, helping users complete tasks more efficiently while reducing learning curves and cognitive load~\cite{ballay1994designing}.

\subsubsection{Experiment Limitations}
Despite the positive feedback from users, it’s important to highlight some experimental constraints. The study was confined to a short session, which doesn’t capture the long-term engagement—spanning months or even years—that is typical for learning and task exploration. Additionally, learning and task exploration are often collaborative processes, suggesting that individual preferences might shape perceptions of specific visualizations. To obtain insights that more accurately reflect the learning and task exploration process, a randomized, controlled, and longitudinal field study would be essential. Such a study, involving users over an extended period, would provide a deeper understanding of how Mindalogue is perceived and integrated into various learning and task exploration workflows, offering a clearer picture of its utility across different scenarios.

\subsection{Application Scenarios and Future Work}
\subsubsection{Expanding to Other Application Areas}
The non-linear interaction design of Mindalogue demonstrates significant advantages in complex information processing fields such as education, research, and knowledge management, particularly in handling interdisciplinary tasks and concept exploration~\cite{card1991information,ballay1994designing}. Qualitative feedback from participants suggests that the system effectively supports knowledge expansion and cross-domain research. Future work could explore its application in fields such as medicine and law, where the demand for highly structured and accurate information is paramount. By integrating more precise domain models and knowledge bases, Mindalogue could become a valuable tool for supporting expert knowledge analysis and reasoning.

\subsubsection{Beyond Node-Based Representations}
Although node-link diagrams (a non-linear topological structure of nodes and links) are the core representation of Mindalogue, the system’s potential is not limited to this format. 
Future iterations could introduce the functionality to establish relationships between two or more nodes, specifically including:

\begin{itemize}
\item \textbf{User-defined relationships}: Users can manually establish custom relationships between two or more nodes based on specific task needs. This provides users with a high degree of flexibility, allowing them to add personalized explanations or annotations according to the task, helping them build their own cognitive framework. For instance, in interdisciplinary tasks, users could create custom relationships to connect seemingly unrelated fields or concepts, forming new cognitive paths~\cite{zhang1994representations}.

\item \textbf{AI-generated relationships}: Based on AI’s understanding of the node content, the system could automatically generate relationships between two or more nodes. This feature would significantly reduce user workload, especially when dealing with large-scale information or complex tasks. AI-generated relationships can help users quickly uncover hidden connections or identify overlooked content~\cite{di2022idea}. For example, in scientific research or multi-dimensional data analysis, the system could automatically detect potential links between different pieces of information, providing users with new insights.
\end{itemize}

These relationships can be established between nodes at the same level or across different levels, with relational lines being generated to illustrate the connections between node contents, further enriching the user’s exploration experience. This multi-node relationship feature not only enhances the system’s flexibility and intelligence but also showcases the adaptability and innovation of non-linear interaction in complex tasks.

\subsection{Summary}
Mindalogue’s non-linear interaction design effectively enhances users' ability to decompose tasks and explore information in complex scenarios, while its graphical representation strengthens users’ understanding of information structures. However, issues with annotation accuracy and the complexity of non-linear interaction remain challenges to be addressed. Future research should continue to explore how external knowledge bases, optimized prompting strategies, and multiple information display methods can be integrated to further improve the system’s performance and adaptability. This study offers new insights into the application of non-linear interaction systems in complex information processing and provides theoretical and practical guidance for future user interaction design.

\section{Conclusion}
In summary, the Mindalogue system takes a significant step forward in enhancing user interaction with LLMs by offering a non-linear graphical interaction mode. Our research demonstrates that the system shows remarkable advantages in complex task decomposition, interdisciplinary exploration, and reducing cognitive load, while flexible graphical content presentation enhances user understanding and operational efficiency. Unlike traditional linear interaction interfaces, Mindalogue allows users to navigate, customize, and analyze information more efficiently, addressing the shortcomings of linear interaction. Looking ahead, this system has broad application potential in fields such as education and research, with further developments enhancing its adaptive capabilities, including AI-generated node relationships. These advances hold promise for driving innovation in how users interact with and comprehend AI-generated content at scale.

\bibliographystyle{ACM-Reference-Format}
\bibliography{main}

\appendix
\section{Prompts}

In this chapter, we provide a detailed exposition of all the prompts utilized in the GPT-Mindalogue prototype environment, which are built upon OpenAI's GPT-4o API. The system and the user are defined as predefined roles that query the API.

\subsection{Generate mind maps}

To ensure that the system generates high-quality mind maps across diverse domains, we categorize the prompts for the system role into two parts: the first part outlines the overall task, aimed at providing the necessary background information for mind map generation; the second part is designed to match the user’s input questions, offering more specific guidance. When a user’s question aligns with the predefined options, the system automatically incorporates relevant prompts, thereby enhancing the relevance and accuracy of the generated content.

\textbf{System(part 1)} You are a useful mind map/undirected graph generating AI that can create a mind map based on any input or instruction. Please generate the mind map according to the following requirements: use \# to denote different levels, including leaf nodes; the number of \# indicates the depth of the level. There can be a maximum of 4 levels. Leaf nodes should have sibling nodes, and ensure that the content of the leaf nodes is more detailed. The total character count of the generated content should be less than 1000 characters. Avoid generating summary statements. The content of the leaf nodes should be complete sentences, with each sentence containing more than 15 characters. The generated content should be logically clear and structurally rigorous.

\textbf{System(part 2)}

If the user's question contains any of the following keywords: Dadaism, Surrealism, Pop Art, French Revolution, Industrial Revolution, Cold War. Then add the following prompt: Please generate an in-depth and creative mind map for {The query provided by the user}.The mind map should explore the historical background, core ideas, key figures, and their profound impact on society, culture, and politics. Delve into the motivations, contradictions, and social responses behind it, and analyze its inheritance and influence in different cultural contexts. Avoid conventional descriptions of events and concepts, and encourage the generation of unique perspectives and depth in content, exploring the implications and influences of the topic on future developments, ensuring the content is novel, rich, and deep.

If the user's question contains any of the following keywords: Machine Learning, Relativity, Quantum Mechanics. Then add the following prompt: Please generate a complex and detailed mind map for {The query provided by the user}.Include the core theories, key algorithms, historical developments, application scenarios, and current technical challenges and future trends. Pay special attention to the latest research progress in the field, and illustrate the application and impact of these technologies with practical cases, avoiding single concept introductions, and encouraging in-depth analysis and discussions on future possibilities.

If the user's question contains any of the following keywords: Marketing Campaign Planning, Brand Promotion, Market Analysis. Then add the following prompt: Please generate a comprehensive and detailed business mind map for {The query provided by the user}. Cover all aspects from market research to brand strategy, advertising, sales channels, and customer experience, and explore the actual effects of different strategies in various market environments. Encourage the generation of flexible and innovative content, avoiding common templated solutions, and provide unique insights through case analysis.

If the user's question contains any of the following keywords: Machine Learning, Relativity, Quantum Mechanics. Then add the following prompt: Please design a creative and inspiring mind map around the following theme: {The query provided by the user}. 1. Explore the background and importance of the theme, using multi-angle and multi-level analytical methods, avoiding a single perspective, emphasizing its future potential and cross-domain relevance. 2. Propose flexible and innovative concepts or design solutions, covering technology, culture, social trends, event planning, and more, encouraging a break from traditional thinking patterns. 3. Describe possible application scenarios, exploring the adaptive value of these ideas in different contexts, and analyze potential opportunities and challenges, avoiding common templates. 4. Construct diverse implementation paths, including the integration of emerging technologies, cultural trends, and market demands, making the proposals forward-looking and widely impactful. Please provide divergent and flexible thinking, ensuring that the content has depth and breadth, capable of inspiring new thoughts and discussions.

If the user's question contains any of the following keywords: bilibili, Xiaohongshu, Forest, Stresswatch, Keep, app, application. Then add the following prompt: Please recommend an application named {The query provided by the user} to others, generating a detailed and flexible mind map around the following aspects, avoiding common templated language, and providing objective and comprehensive information: 1. Application Functions: Explore the core functions of the application and its uniqueness, and propose potential innovative applications or functional expansions. 2. User Experience: Analyze the interface design, user interaction, and usage experience from diverse user perspectives, discussing how to further enhance usability and user satisfaction. 3. Applicable Scenarios: Step outside conventional thinking and explore the potential value of the application in non-traditional or unexpected scenarios, sharing flexible usage methods. 4. Market Positioning: Conduct an in-depth analysis of the application's competitiveness in the existing market, proposing innovative market positioning strategies or niche market opportunities. 5. Personal Recommendation Reasons: Share personal experiences and insights, providing reasons for recommending the application, and offering unique perspectives or usage suggestions to inspire others. 6. Future Prospects: Explore the possible development directions and innovative opportunities for the application, proposing creative ideas for adapting to future user needs.

If the user's question contains any of the following keywords: city, municipality, district, county, town. Then add the following prompt: Please provide a detailed description of a city, which is {The query provided by the user}. Please generate a mind map from the following aspects and provide objective and comprehensive information: 1. City Characteristics: Introduce the geographical location, climate, architectural style, and unique features of the city in detail, considering the city's performance in different seasons or times. 2. Culture and Lifestyle: Delve into the cultural atmosphere, social customs, major festivals, and lifestyles of the city, analyzing its historical background and cultural diversity. 3. City Economy: Analyze the city's economic development status, main industries, and commercial vitality, exploring the city's role and position in the global or regional economy. 4. Personal Experiences: Share your feelings, memories, or unique experiences about the city based on personal experiences, and discuss how these experiences shape your impression of the city. 5. Future Prospects: Explore the future development potential of the city from multiple angles, including urban planning, sustainable development, and social change, proposing your expectations or suggestions for the city. Please avoid using common templated language, encourage the generation of innovative and flexible content, and ensure the depth and breadth of the information.

In other cases, add the following prompt: Please generate a unique and profound mind map for {The query provided by the user}. Ensure that it covers the core elements of the topic, digging deep into potential connections and complexities, avoiding templated language and repetitive content, striving to generate inspiring and innovative content.

\textbf{User} [The query provided by the user.]

\subsection{Node explanation}

\textbf{System} I want to explain a certain node in the mind map, where the parent node information is: \{parent\_information\}, and the node information is: \{node\_information\}. Please use the parent node information as background information to explain the selected node. The explanation should be logically clear, rich in content, and not repeat the parent node information. The generated answer should start with \#, and the content should be a paragraph of 2 to 3 sentences, serving as the child node of the selected node. Do not generate summary statements before or after the explanation. Only respond with the explanation content, without splitting levels, and keep the total character limit within 100 characters.

\subsection{Node examples}

\textbf{System} Please use the following parent node information as background: \{parent\_information\} to generate at least 1 and at most 3 relevant cases for the selected node \{node\_information\}. Each case should be a paragraph and serve as a child node of the selected node, which must be indicated by one additional \# compared to the selected node \{node\_information\}. Do not generate summary statements before or after the explanation. The language should be logically clear, with a repetition rate not exceeding 50\% compared to the child node information, and each case should be independent and highly relevant to the node information.

\subsection{Custom questions}

\textbf{System} I want to ask a question about a specific node in the mind map, with the parent node information: \{parent\_information\} and the selected node information: \{node\_information\}. Please use the parent node information as background and answer the following question regarding the selected node: \{The query provided by the user\}. The generated content should be a paragraph consisting of 2 to 3 sentences, maintaining clear logical language and rich content, and the answer should not repeat the parent node and selected node information. The generated answer should start with \# and the total character limit for the answer is 100 characters.

\textbf{User} [The query provided by the user.]

\end{document}